\documentclass[12pt,cite,a4paper]{article}

\usepackage{amsmath}
\usepackage{amssymb}
\usepackage{a4wide}
\usepackage{amssymb}
\usepackage{graphicx}

\newcommand{\cA}{\mathcal{A}}
\newcommand{\cB}{\mathcal{B}}

\numberwithin{equation}{section}

\begin{document}
\rightline{PUPT-2268}
\rightline{QMUL-PH-08-10}
\vspace{2truecm}

\centerline{\Large \bf $\mathcal{N} = 8$ superconformal gauge theories and M2 branes}

\vspace{1.3truecm}

\centerline{
    { Sergio Benvenuti${}^{a,}$}\footnote{sbenvenu@princeton.edu},
    {  Diego Rodr\'{\i}guez-G\'omez${}^{a,b,}$}\footnote{drodrigu@princeton.edu},
    {  Erik Tonni${}^{c,}$}\footnote{e.tonni@df.unipi.it}
    {and}
    {  Herman Verlinde${}^{a,}$}\footnote{verlinde@princeton.edu}}
\vspace{.8cm}
\centerline{{\it ${}^a$Department of Physics, 
Princeton University}}
\centerline{{\it Princeton, NJ 08544, USA}}
\vspace{.4cm}
\centerline{{\it ${}^b$ Center for Research in String Theory, Queen Mary University of
    London}} \centerline{{\it Mile End Road, London, E1 4NS, UK}}
\vspace{.4cm}
\centerline{{\it ${}^c$ Dipartmento di Fisica, 
Universit\`a di Pisa and INFN, sezione di Pisa,}}  
\centerline{{\it Largo Bruno Pontecorvo 3, 56127 Pisa, Italy}}

\vspace{2.5truecm}

\centerline{\bf ABSTRACT}
\vspace{.5truecm}

Based on recent developments, in this letter we find $2+1$ dimensional gauge theories with scale invariance and $\mathcal{N} = 8$ supersymmetry. The gauge theories are defined by a lagrangian and are based on an infinite set of $3$-algebras, constructed as an extension of ordinary Lie algebras. Recent no-go theorems on the existence of $3$-algebras are circumvented by relaxing the assumption that the invariant metric is positive definite. The gauge group is non compact, and its maximally compact subgroup can be chosen to be any ordinary Lie group, under which the matter fields are adjoints or singlets.
The theories are parity invariant and do not admit any tunable coupling constant. In the case of $SU(N)$ the moduli space of vacua contains a branch of the form $(\mathbb{R}^8)^N/ S_N$. These properties are expected for the field theory living on a stack of M2 branes.

\noindent

\newpage

\section{Introduction}

In a series of papers by Bagger and Lambert \cite{Bagger:2006sk,Bagger:2007jr, Bagger:2007vi}  and Gustavsson \cite{Gustavsson:2007vu, Gustavsson:2008dy}, a new set of $2+1$ dimensional theories theories  enjoying $\mathcal{N}=8$ supersymmetry, $SO(8)$ global symmetry and scale invariance has been proposed. These theories are supposed to describe multiple coincident M2 branes; a Chern-Simons term and a sextic potential for the scalars are present in the lagrangian, as expected \cite{Schwarz:2004yj}. This framework involves an unusual algebraic structure called $3$-algebra: a vector space endowed with a set of structure constants with four indices $f^{abc}\,_d$. The structure constants satisfy antisymmetry in the upper three indices and a fundamental identity analogue to the Jacobi identity satisfied by the structure constants $f^{\alpha \beta}\,_{\gamma}$ of ordinary Lie algebras.

The requirement that the 3-algebra has a positive definite metric is very strong:  it was recently proven in \cite{Papadopoulos:2008sk, Gauntlett:2008uf} that there is only one such $3$-algebra, called $\mathcal{A}_4$ (or linear sums thereof). $\mathcal{A}_4$ is $4$-dimensional and $f^{abcd} = \epsilon^{abcd}$. The corresponding $\mathcal{N}=8$ theory has gauge symmetry $SO(4)$ and is expected to describe two M2 branes sitting at the origin of $\mathbb{R}^8/\mathbb{Z}_2$ \cite{VanRaamsdonk:2008ft, Lambert:2008et, Distler:2008mk}. It was shown in \cite{Mukhi:2008ux} that, upon giving a VEV to a scalar, it is possible to recover the theory on coincident D2 branes via a novel type of Higgs mechanism. For additional results see \cite{Bandres:2008vf, Berman:2008be, Morozov:2008cb, Gran:2008vi,  Ho:2008bn, Gomis:2008cv, Bergshoeff:2008cz, Hosomichi:2008qk, Ho:2008nn}.

The problem of describing $N$ M2 branes in flat space is the main objective of this letter.

It seems reasonable to relax some of the constraints in order to evade the no-go theorems, and maybe describe $N$ M2's. This direction has been pursued in \cite{Morozov:2008cb, Gran:2008vi}. In \cite{Gran:2008vi} Gran, Nilsson and Petersson proposed to focus only on the equations of motion, relying on the fact that many $3$-algebras exist if we don't require the existence of a metric. It's important to remark that in the Bagger-Lambert work at the level of the equation of motion the metric is not used. The metric is needed in order to have a Lagrangian and gauge invariant local operators, such as the energy-momentun tensor.

\vspace{0.2cm}

In this letter we relax the assumption that the metric on the $3$-algebra is positive definite. This allows us to find an infinite set of $3$-algebras. The construction starts from any ordinary Lie algebra $\mathcal{G}$ and consists in adding two directions to $\mathcal{G}$ as a vector space, which we call ${+}$ and $-$, thus the $3$-algebra has dimension $dim(\mathcal{G})+2$. Using indices $a,b, \ldots = \{+, -, \alpha\}$, the structure constant are given in terms of the $\mathcal{G}$-structure constants $f^{\alpha \beta}\,_{\gamma}$ as
\begin{equation}\label{basic3A}
f^{+ \alpha \beta}\,_{\gamma} = - f^{\alpha + \beta}\,_{\gamma} = f^{\alpha \beta +}\,_{\gamma} = f^{\alpha \beta}\,_{\gamma} \hspace{2cm} f^{\alpha \beta \gamma}\,_{-} = f^{\alpha \beta \gamma} 
\end{equation}
all other possible components of $f^{a b c}\,_d$ simply vanish. This $dim(\mathcal{G})+2$ dimensional $3$-algebra is related to a similar $dim(\mathcal{G})+1$ algebra proposed in \cite{Gustavsson:2008dy, Gran:2008vi, Awata:1999dz}.\footnote{More precisely if we project (\ref{basic3A}) along the $dim(\mathcal{G})+1$ dimensional subspace generated by $\{ + , \alpha \}$ we obtain the $3$-algebra discussed in \cite{Gustavsson:2008dy, Gran:2008vi, Awata:1999dz}. We can also project along the subspace generated by $\{ - , \alpha \}$.} One nice feature of (\ref{basic3A}) is the existence of an invariant metric, given in terms of the standard  metric on $\mathcal{G}$ 
\begin{equation}\label{trace}
\left(\begin{array}{cc|ccc}
0 & -1 & 0 & \dots & 0\\
-1 & 0 & 0 & \dots & 0\\ \hline
0 & 0 &  && \\
\vdots & \vdots & & h_{\mathcal{G}}&\\
0 & 0 &  &&
\end{array}\right)
\end{equation}
This metric is clearly non positive definite, having signature $(dim(\mathcal{G})+1, 1)$ if $\mathcal{G}$ is compact and semisimple. The metric is invariant under the symmetry group of the equations of motion, which turns out to be a Inonu-Wigner contraction of $\mathcal{G}\otimes \mathcal{G}$. Our theory has thus a non compact gauge group of dimension $2 \, dim(\mathcal{G)}$, that can be embedded in $SO(dim(\mathcal{G})+1, 1)$. The $dim(\mathcal{G})+2$ scalar fields $X^I$ and fermion fields $\Psi$ transform in the $8$ of $SO(8)$. Two fields ($X^I_\pm$ and $\Psi_\pm$) are singlets of $\mathcal{G}$ (the maximally compact subgroup of the full non compact gauge group), while the other fields $X^I_\alpha, \Psi_\alpha$ transform in the adjoint representation. Since the gauge theory discussed in this paper can be recast in the Bagger-Lambert framework, it automatically enjoys $\mathcal{N}=8$ supersymmetry. 

One new feature is that the gauge interactions turns out to be of the BF-type. BF theories do not admit a tunable coupling constant, and this property extends to the full superconformal theory. This is expected for the gauge theory living on M2 branes. Also parity invariance is preserved.

\vspace{0.2cm}
In the main body of this letter, section \ref{main example}, we discuss in detail the classical aspects of the theory based on the $3$-algebra (\ref{basic3A}), finding the gauge symmetries and an explicit lagrangian. We show how the overall coupling in front of the lagrangian can be reabsorbed via rescaling of the fields. We also analyze the moduli space of vacua and the mass of the low energy fluctuations. In the case of $\mathcal{G} = SU(N)$, the moduli space of vacua contains a branch of the form $\frac{(\mathbb{R}^8)^{N}}{S_N}$, as expected for a theory describing $N$ M2 branes in flat $11$-dimensional space. We conclude in section \ref{concl} with some speculative remarks about the quantization and the unitarity of the theory, and we comment on the possible relation to $M$ theory. 

\vspace{0.5cm}
\textbf{Note added}: after this letter was completed, the preprint \cite{gomis} appeared on the arXiv with substantial overlap with our results.

\section{New $\mathcal{N}=8$ superconformal gauge theories}\label{main example}

\subsection{Mini-review of the Bagger-Lambert framework}\label{review}

The theory on coincident M2 branes should involve 8 real scalar fields $X^I_a$, $I=1,\dots,8$. Then, $\mathcal{N}=8$ supersymmetry requires a 16 component spinor $\Psi^I_a$, which we can take a chiral spinor of $SO(8)$. The fields carry an internal index $a$ running from $1$ to $D$, where $D$ is the dimension of the $3$-algebra. With these ingredients, Bagger and Lambert proposed the following $\mathcal{N}=8$ SUSY transformations, consistent with classical scale invariance:
\begin{eqnarray}
\label{SUSY}\nonumber
&&\delta X^I_a=i\bar{\epsilon}\Gamma^I\Psi_a\ ,\\
&&\delta\Psi_a=D_{\mu}X^I_a\Gamma^{\mu}\Gamma_I\epsilon-\frac{1}{6}X^I_bX^J_cX^K_df^{bcd}\,_a\Gamma_{IJK}\epsilon\ ,\\
&& \delta (\tilde{A}_{\mu})^a_b=i\bar{\epsilon}\Gamma_{\mu} \Gamma_I X^I_c \Psi_d f^{cda}\,_b\ ;\nonumber
\end{eqnarray}
where $f^{abc}\,_d$ are the structure constants of the 3-algebra and are completely antisymmetric in the $3$ upper indices.
The closure of the SUSY transformations implies the equations of motion 
\begin{eqnarray}
\label{eomg}\nonumber
&& \Gamma^{\mu}D_{\mu}\Psi_a+\frac{1}{2}\Gamma_{IJ}X^I_cX^J_d\Psi_b f^{cdb}\,_a=0\ ,\\
&& D^2X^I_a-\frac{i}{2}\bar{\Psi}_c\Gamma^I_JX^J_d\Psi_bf^{cdb}\,_a+\frac{1}{2}f^{bcd}\,_af^{efg}\,_d\, X^J_bX^K_cX^I_eX^J_fX^K_g=0\ ,\\
&& (\tilde{F}_{\mu\nu})^b_a+\epsilon_{\mu\nu\lambda}\big(X^J_cD^{\lambda}X^J_d+\frac{i}{2}\bar{\Psi}_c\Gamma^{\lambda}\Psi_d\big)f^{cdb}\,_a=0\ .\nonumber
\end{eqnarray}
The fields transform in the following way under gauge transformation
\begin{equation}
\label{gaugetransf}
\delta X^I_a=\tilde{\Lambda}^b_aX^I\ , \qquad \delta \Psi_a=\tilde{\Lambda}^b_a\Psi_ b\ , \qquad \delta(\tilde A_{\mu})^b_a=
D_{\mu}\tilde{\Lambda}^b_a\ .
\end{equation}
where  $\tilde{\Lambda}^a_{b} = \Lambda_{mn} f^{mna}\,_b$ and $(\tilde{A}_\mu)^a_{b} = (A_\mu)_{mn} f^{mna}\,_b$.
The gauge group is generated by the $\tilde{\Lambda}^a_{b}$, while the antisymmetric $\Lambda_{mn}$ are auxiliary parameters. The gauge group is thus a subgroup of $GL(D)$. (If we add a metric of signature $(D - k , k)$ on the $3$-algebra, then we can say that the gauge group is a subgroup of $SO( D - k, k)$). In order for the equations of motion to be consistent with gauge symmetry and supersymmetry, a constraint on the structure constants has to be satisfied:
\begin{equation}
\label{FI}
f^{efg}\,_d f^{abc}\,_g-f^{efa}\,_gf^{bcg}\,_d-f^{efb}\,_gf^{cag}\,_d-f^{efc}\,_gf^{abg}\,_d=0\ .
\end{equation}
This is known as the fundamental identity. At the level of equations of motion the only constraints on the structure constants come from the fundamental identity and antisymmetry in the upper 3 indices.

\subsection{The special case of the $dim(\mathcal{G})+2$ $3$-algebra (\ref{basic3A})}
As a first step we determine the gauge group for the special choice of structure constants (\ref{basic3A})
\begin{equation}
f^{+ \alpha \beta}\,_{\gamma} = - f^{\alpha + \beta}\,_{\gamma} = f^{\alpha \beta +}\,_{\gamma} = f^{\alpha \beta}\,_{\gamma} \hspace{2cm} f^{\alpha \beta \gamma}\,_{-} = f^{\alpha \beta \gamma} 
\end{equation}

Since the fields $X_\alpha$ transform in the adjoint of $\mathcal{G}$ we can introduce the matrices $T^\alpha$ such that
\begin{eqnarray}
&& X^I = X^I_\alpha T^\alpha \\
&& [T^\alpha , T^\beta] = f^{\alpha \beta}\,_{\gamma} \, T^{\gamma} \\
&& {\rm Tr}\Big(T^\alpha T^\beta\Big) = \delta^{\alpha \beta}
\end{eqnarray}

We focus on the equations of motion (\ref{eomg}) but, for the sake of simplicity, we set to zero the fermions and the gauge fields. (\ref{eomg}) becomes
\begin{eqnarray}
\label{eomsc}\nonumber
&& \partial^2 X_+^I = 0\ ,\\ 
&& D^2 X_-^I =   \frac{1}{2} X^I_+  {\rm Tr}\Big( [X^J, X^K]^2 \Big) - X^J_+ {\rm Tr}\Big( [X^J, X^K]  [X^I, X^K] \Big) \\
&& \nonumber
D^2X^I  =  (X_+^J)^2[X^K,[X^I,X^K]] - X_+^J X_+^I[X^K,[X^J,X^K]] - X_+^JX_+^K [X^K, [X^I, X^J]]\ .
\end{eqnarray}

Where for the moment the precise definition of the covariant derivative $D^\mu$ is not important. We now want to study the symmetries of (\ref{eomsc}) under global transformations of the fields $(X^I_\pm, X^I)$. The symmetry generators act on the matter field through the matrix 
\begin{equation}
\tilde{\Lambda}^a_{b} = \Lambda_{mn} f^{mna}\,_b\,\,.
\end{equation}
 From (\ref{basic3A}) it is easy to see that 
 \begin{equation}
 \tilde{\Lambda}^-_{+} = \tilde{\Lambda}^\alpha_{+} = \tilde{\Lambda}^+_{-} = 0
\end{equation}
 while $\tilde{\Lambda}^+_{\alpha}$ and $\tilde{\Lambda}^{\alpha}_-$ are given in terms of $\Lambda_{\alpha \beta}$. These transformations are in one-to-one correspondence with the group $\mathcal{G}$. These are obviously symmetries of (\ref{eomsc}), denoted by $T_c^\alpha$.
 
 An additional set of $dim(\mathcal{G})$ symmetries are however present, corresponding to $\tilde{\Lambda}^\alpha_{\beta}$, which comes from $\Lambda_{\alpha +}$. These additional symmetries, which we denote by $T_{nc}^\alpha$, act on the fields as
 \begin{eqnarray}\nonumber
 \delta X^I_+ &=& 0 \\
 \delta X^I_- &=&   {\rm Tr} \Big( M X^I \Big) + \frac{1}{2} {\rm Tr}\Big( M^2 \Big) X^I_+ \\ 
 \delta X^I     &=&    M X^I_+ \nonumber
 \end{eqnarray}
 Where $M$ is a matrix in the adjoint of $\mathcal{G}$ which is not necessarily infinitesimal. It is clear that at the infinitesimal level the transformations $T_{nc}$ commute among themselves.
 
 In order to find the full set of commutation relations it is useful to write the generators of the symmetry group as $(dim(\mathcal{G})+2) \times (dim(\mathcal{G}) + 2)$ matrices.
 
 \begin{equation}\label{symmgen}
T_c^\alpha  =  \left(\begin{array}{cc|ccc}
0 & 0 & 0 & \dots & 0\\
0 & 0 & 0 & \dots & 0\\ \hline
0 & 0 &  && \\
\vdots &  \vdots & & T^{\alpha}&\\
0 & 0 &  &&
\end{array}\right)
\hspace{2.3cm}
T_{nc}^\alpha  = \hspace{-0.3cm}  \begin{array}{cc}
& 
\begin{array}{ccccccccc}
&  &  &  & \hspace{0cm}\alpha\; &  & \\
 &  &  &  & \hspace{-0cm}\downarrow \;&  &
\end{array}
\\
\begin{array}{c}
\\  \\  \\  \\ \alpha\hspace{-0.05cm}\rightarrow \hspace{-0.15cm}\\ \vspace{.1cm}\\ \\
\end{array}
&
\hspace{-.45cm}
\left(
\begin{array}{cc|ccccccc}
0 & 0 & 0 & \dots & 0 & \dots & 0\\
0 & 0 & 0 & \dots & 1 & \dots & 0\\
\hline 
0 & 0 &  &  &  &  & \\
\vdots & \vdots &  &  &  &  & \\
1 & 0 &  &  &  0 &  & \\
\vdots & \vdots  &  &  &  &  & \\
0 & 0 &  &  &  &  & \\
\end{array}
\right)
\end{array}
\end{equation}

It is possible to check that the algebra is
\begin{equation}
\label{algebra}
[T_c^{\alpha},T_c^{\beta}]=  f^{\alpha\beta}\,_{\gamma}T_c^{\gamma}\ ,\qquad [ T_{nc}^{\alpha}, T_{nc}^{\beta}]=0\ ,\qquad [ T_{nc}^{\alpha},T_c^{\beta}]= f^{\alpha\beta}\,_{\gamma} T_{nc}^{\gamma}\ .
\end{equation}

To gain further insight we modify this algebra with a parameter $\epsilon$
\begin{equation}
\label{algebrad}
[T_c^{\alpha},T_c^{\beta}]=  f^{\alpha\beta}\,_{\gamma}T_c^{\gamma}\ ,\qquad [ T_{nc}^{\alpha}, T_{nc}^{\beta}]= \epsilon f^{\alpha\beta}\,_{\gamma}T_c^{\gamma}\ ,\qquad [ T_{nc}^{\alpha},T_c^{\beta}]= f^{\alpha\beta}\,_{\gamma} T_{nc}^{\gamma}\ .
\end{equation}
so that in the limit $\epsilon\rightarrow 0$ we recover (\ref{algebra}). It is easy to see \footnote{We can now re-scale the generators and combine them into two sets $\{T_\pm^{\alpha}\}$ in such a way that the deformed algebra is 
$[T_+^{\alpha} , T_-^{\beta}]=0\ ,\quad [T_+^{\alpha} , T_+^{\beta}]= f^{\alpha\beta}\,_{\gamma} T_+^{\gamma}\ ,\quad [ T_-^{\alpha} , T_-^{\beta}] = f^{\alpha\beta}\,_{\gamma} T_-^{\gamma}$}
that (\ref{algebrad}) is the Lie algebra of $\mathcal{G} \otimes \mathcal{G}$. The symmetry algebra of the ungauged theory is thus a non compact Inonu-Wigner contraction of $\mathcal{G} \times \mathcal{G}$. This analysis has been carried out just at the level of scalar fields, but it's easy to see that it extends to the full set of equation of motion. Notice that the theory classically have a shift symmetry acting on the $8$ scalars $X^I_-$. The role of this symmetry is not clear.

The theory has thus $2\, dim\mathcal{G}$ gauge fields, we denote $\cA^\mu$ and $\cB^\mu$ the fields associated the compact part $\mathcal{G}$ and the non compact part, respectively:
\begin{equation}
 (\cA_\mu)_\alpha = (A_\mu)_{+ \alpha}  \hspace{2 cm}  (\cB_\mu)_\alpha = (A_\mu)_{\beta \gamma} f^{\beta \gamma}\,_\alpha 
\end{equation}
and $\cA_\mu = (\cA_\mu)_\alpha T^\alpha$, $\cB_\mu = (\cB_\mu)_\alpha T^\alpha$. So $\cA_\mu$ and $\cB_\mu$ are matrices in the adjoint of $\mathcal{G}$.
The covariant derivatives are defined as
\begin{eqnarray}
D_{\mu}X^I&=&\partial_{\mu}X^I-2 [ \cA_{\mu} , X^I ] - \cB_{\mu}X_ +^I\ ;\\
D_{\mu}X^I_-&=&\partial_{\mu}X_-^I-{\rm Tr} \Big( \cB_{\mu}X^I \Big)\ ;\\
D_{\mu}X^I_+&=&\partial_{\mu}X^I_+\ ,
\end{eqnarray}
and similarly for the fermions. $X^I,\,\Psi$ are fields in the adjoint of $\mathcal{G}$ and $X_\pm^I, ,\Psi_\pm$ are singlets. The gauge parameters can also be assembled in matrices of the adjoint of $\mathcal{G}$ as
\begin{equation}
\Lambda = \Lambda_{+ \alpha} T^\alpha \hspace{1.5cm} M = \Lambda_{\beta \gamma} f^{\beta \gamma}\,_\alpha T^\alpha \,,
\end{equation}
under which the gauge fields transform as
\begin{eqnarray}
\delta \cA_\mu & = & \partial_\mu \Lambda - 2 [ \cA_\mu , \Lambda ] \\
\delta \cB_\mu & = & \partial_\mu M - 2 [ \cA_\mu , M ] - 2 [ \cB_\mu , \Lambda ] 
\end{eqnarray}
The equations of motion for the gauge fields, in absence of matter fields, are simply $(\tilde{F}_{\mu\nu})^b_a = 0$. The non trivial ones are 
\begin{equation}\label{gaugeeq}
(\tilde{F}_{\mu\nu})^+_\alpha = 0 \hspace{2 cm} (\tilde{F}_{\mu\nu})^\beta_\alpha = 0 \,, 
\end{equation}
After contracting with $\epsilon^{\lambda \mu \nu}$, in terms of $\cA_\mu$ and $\cB_\mu$ (\ref{gaugeeq}) imply
\begin{eqnarray}
 0 & = & \epsilon^{\lambda \mu \nu } ( \partial_\mu \cB_\nu - [ \cA_\mu , \cB_\nu ] )  \\
 0 & = & \epsilon^{\lambda \mu \nu} ( \partial_\mu \cA_\nu - [ \cA_\mu, \cA_\nu] ) 
\end{eqnarray}

The action for the gauge fields $\cA$ and $\cB$, from which the above equations can be derived, is of the $BF$-type:
\begin{equation}\label{LAGcs}
S_{gauge}  =
\int  d^3x \, \, \epsilon^{\lambda\mu\nu} \,{\rm Tr}\Big( \cB_{\lambda} (\partial_{\mu}\cA_{\nu} - [\cA_{\mu}, \cA_{\nu}]) \Big)
 \end{equation}

It is clear that this action is parity invariant, if we define parity to act on the gauge field $\cB^\mu$ by flipping its sign. Notice the no tunable coupling constant can appear in (\ref{LAGcs}).

\subsection{The Lagrangian}
The metric associated to our choice of structure constants is
\begin{displaymath}
h^{a b} = \left(\begin{array}{cc|ccc}
0 & -1 & 0 & 0 & 0\\
-1 & 0 & 0 & 0 & 0\\ \hline
0 & 0 &  && \\
0 &  0 & & \mathcal{I}_{\mathcal{G}}&\\
0 & 0 &  &&
\end{array}\right)
\end{displaymath}
It is easy to check that this metric is invariant under the transformations (\ref{symmgen}). Using this metric it is possible to write down a leomscagrangian. As a first illustrative step we consider just the scalar part. The equations of motion (\ref{eomsc}) are derived from the following ungauged lagrangian:
\begin{equation}
\label{LS}
\mathcal{L}_S = -\frac{1}{2}{\rm Tr}\Big(\partial_{\mu}X^I \partial_{\mu}X^I\Big) + \partial_{\mu}X^I_+\partial_{\mu}X_-^I - \frac{1}{12} {\rm Tr}\Big(X_+^I [ X^J, X^K] + X^J_+ [ X^K, X^I ] + X_+^K [ X^I , X^J ]\Big)^2 
\end{equation}
It is important that this lagrangian does not have a coupling constant. Indeed, the same is true for the gauge lagrangian (\ref{LAGcs}). \footnote{This $3$D theory is of the form $\partial X_+ \partial X_-  - (\partial X)^2 + X_+^2 X^4$, which is scale invariant and does not admit a tunable parameter but still seems non trivial. The analogous scale invariant lagrangian in $4$D, $\partial X_+ \partial X_- - (\partial X)^2 + X_+^2 X^2$, is instead much simpler, since the equations of motions are all linear if solved in the right order: first the one coming from $X_-$ ($\partial^2 X_+ = 0$), then the one coming from $X$ ($\partial^2 X = X_+^2 X$) and finally the equation coming from $X_+$ ($\partial^2 X_- = - 2 X_+ X^2$).} 

We finally consider the complete gauged lagrangian, with $\mathcal{N}=8$ SUSY, $SO(8)$ global symmetry, from which the equations of motion follow.\footnote{Notice that in \cite{Bagger:2007jr} the lagrangian was derived by the eqs. of motion under the assumption that the metric is positive definite, so we cannot use directly their results. At the end however we get the same form of the lagrangian, defining $f^{abcd} = h^{de} f^{abc}\,_e$.} The full lagrangian contains standard kinetic terms with covariant derivatives, the gauge term (\ref{LAGcs}), a sextic potential and Yukawa couplings:
\begin{eqnarray}\label{LAGa}
\mathcal{L}&=&-\frac{1}{2} h^{a b} D_{\mu}X^I_a D_{\mu}X^I_b +\frac{i}{2} h^{a b} \bar{\Psi}_a\Gamma^{\mu}D_{\mu}\Psi_b +\epsilon^{\mu\nu\lambda} {\rm Tr}\Big( \cB_{\lambda} (\partial_{\mu}\cA_{\nu} - [\cA_{\mu}, \cA_{\nu}]) \Big) \nonumber \\ 
&& + \frac{1}{12} h^{m n} f^{abc}\,_m f^{efg}\,_n X_a^I X_ b^J X^K_c  X_e^I X_ f^J X^K_g +\frac{i}{4}  h^{d e}  f^{abc}\,_e X^I_a X^J_b \bar{\Psi_c}\Gamma_{IJ} \Psi_d
\end{eqnarray}
In this form the invariance under the non compact gauge group is manifest, moreover from the results of Bagger and Lambert it is clear that (\ref{LAGa}) is $\mathcal{N}=8$ supersymmetric.

We can rewrite the lagrangian in a $\mathcal{G}$-invariant notation:
\begin{eqnarray}\mathcal{L}&=&-\frac{1}{2}{\rm Tr}\Big(D_{\mu}X^ID_{\mu}X^I\Big)+D_{\mu}X^I_+D_{\mu}X_-^I+\frac{i}{2}{\rm Tr}\Big(\bar{\Psi}\Gamma^{\mu}D_{\mu}\Psi\Big)-\frac{i}{2}\bar{\Psi}_+\Gamma^{\mu}D_{\mu}\Psi_--\frac{i}{2}\bar{\Psi}_-\Gamma^{\mu}D_{\mu}\Psi_+ \nonumber \\ \nonumber
&&+ \epsilon^{\mu\nu\lambda} {\rm Tr}\Big( \cB_{\lambda} (\partial_{\mu}\cA_{\nu} - [\cA_{\mu}, \cA_{\nu}]) \Big) - \frac{1}{12} {\rm Tr}\Big(X_+^I [ X^J, X^K] + X^J_+ [ X^K , X^I ] + X_+^K [ X^I , X^J ]\Big)^2 \label{THELAG}
\nonumber\\ 
&& +\frac{i}{2}{\rm Tr}\Big(\bar{\Psi}\Gamma_{IJ}X_ +^I[X^J,\Psi]\Big)+\frac{i}{4}{\rm Tr}\Big(\bar{\Psi}\Gamma_{IJ}[X^I,X^J]\Psi_+\Big)-\frac{i}{4}{\rm Tr}\Big(\bar{\Psi}_+\Gamma_{IJ}[X^I,X^J]\Psi\Big)\ ,\end{eqnarray}
$3$-dimensional parity is preserved if the fields $\cB_\mu$, $X^I$ and $\Psi$ are parity-odd. 

\subsection{Absence of coupling constant}

As for the previously studied truncations, the lagrangian (\ref{THELAG}) does not admit any tunable coupling constant. Indeed, had we considered including a coupling constant $\frac{1}{g^2}\mathcal{L}(X_{\pm},X,\mathcal{B},\mathcal{A})$, we could redefine 
\begin{equation}
X^I=g\, Y^I\, \qquad X^I_+=g^{-1}\,Y_+^I\,\qquad X_-^I=g^3\, Y_-^I\,\qquad \mathcal{B}=g^2\,\tilde{\mathcal{B}},\,
\end{equation}
(and consistently the same for the fermions) in such a way that we have 
\begin{equation}
\frac{1}{g^2}\mathcal{L}(X_{\pm},X,\mathcal{B},\mathcal{A})=\mathcal{L}(Y_{\pm},Y,\tilde{\mathcal{B}},\mathcal{A})\, ,
\end{equation}
\textit{i.e.} the coupling constant can be always reabsorbed. This is a highly non-trivial hint that (\ref{THELAG}) is related to M2 branes. 


\subsection{Comments on the physical spectrum}

Given that the metric in (\ref{THELAG}) is not positive definite, one could worry about the presence of negative norm states in the quantum theory. Since this problem is already present in the ungauged theory, we can start considering just (\ref{LS}). The field $X_-^I$ appears only through $\partial X_+^I\partial X_-^I$, we could perform the functional integral over it, which would lead to a functional delta localizing the $X_+^I$ integral on the solutions to $\partial^2X_ +^I=0$ (this is the quantum-mechanical counterpart of the classical observation that we can regard $X_-^I$ as a Lagrange multiplier, thus enforcing a constraint). We would be left with an effective theory whose partition function is
\begin{eqnarray}
\label{Zeff}
\mathcal{Z}&=&\int \mathcal{D}X_+^I\,\mathcal{D}X^I\,\delta(\partial^2X_+^I)\,\\ \nonumber && {\rm exp}\Big\{i\int -\frac{1}{2}{\rm Tr}\Big(\partial_{\mu}X^I \partial_{\mu}X^I\Big) - \frac{1}{12} {\rm Tr}\Big(X_+^I [ X^J, X^K] + X^J_+ [ X^K, X^I ] + X_+^K [ X^I , X^J ]\Big)^2 \Big\}\ .
\end{eqnarray}
This theory can be regarded as a version of $\lambda \phi^4$ theory in 3 dimensions where we integrate over all the possible $\lambda$ (which in general are space-time dependent, since they will be harmonic functions in 3 dimensions with suitable boundary conditions). However, the coupling $\lambda$ is schematically $(X_+^I)^2$, which ensures that the theory (\ref{Zeff}) does not contain any negative norm states. Since the $X_-^I$ integral is exact, we believe that this hints that the negative norm states can be consistently decoupled from the physical Hilbert space.\footnote{The constraint $\partial^2X_+$ imposed by the delta function in (\ref{Zeff}) can be regarded as the condition that the current associated to the $X_-$ shift symmetry is divergence-free. Therefore, operators that act within the physical Hilbert space should be invariant under the shift symmetry.} We have not performed the analysis in the fully gauged lagrangian (\ref{THELAG}), but we expect that, along the same lines as in the ungauged case, it should be possible to consistently decouple negative norm states even-though the presence of a gauge field (the $\mathcal{B}$ field) of a non-compact gauge symmetry.

One important aspect of having a metric on the space of fields invariant under the symmetry transformations is that it's possible to construct local gauge invariant observables as
\begin{equation}
\mathcal{O}^{IJ}(x) = h^{a b} X^I_a(x) X^J_b(x) = {\rm Tr} \Big( X^I X^J \Big) - X^I_- X^J_+ - X^J_- X^I_+
\end{equation}
This scalar operator transforms in the $35 \oplus 1$ of $SO(8)$. The $35$ should be BPS and should have a dual in the graviton supermultiplet of $11$-dimensional supergravity reduced on $AdS_4 \times S^7$. We can proceed to construct higher order operators as $X^{I_1}_+ \ldots X^{I_n}_+ \mathcal{O}^{IJ}$ and decompose them into $SO(8)$ irreps. 

Note however that this class operators are not invariant under the $X_-$ shift symmetry. Deforming the lagrangian by adding these operators would translate in a different constraint of the form $\partial^2X_+=f(X_+)$. Therefore, it remains to be seen whether this operators survive as physical operators after the negative norm states have been eliminated.

Similarly there are the $SO(8)$ currents:
\begin{equation}
\mathcal{K}_\mu^{IJ}(x) = h^{a b} ( X^I_a(x) \partial_\mu X^J_b(x) -  X^J_a(x) \partial_\mu X^I_b(x) + fermions)
\end{equation}
The energy momentum tensor should have a very similar structure.

\subsection{Moduli space of vacua for $\mathcal{G} = SU(N)$}
In this section we perform a preliminary study of the moduli space of vacua of the theory.

Setting the fermions and the gauge fields to zero, and the scalar fields to be constant, we can look for the moduli space of supersymmetric vacua as those configurations preserving supersymmetry. This requires that
\begin{equation}
X^I_cX^J_dX^K_bf^{cdb}\,_a=0\ .
\end{equation}

In our case these equations are equivalent to
\begin{eqnarray}
\label{vacua}
0 & = & X_+^I [ X^J, X^K] + X^J_+ [ X^K , X^I ] + X_+^K [ X^I , X^J ]\ ,\\
0 & = &  {\rm Tr}\Big( X^I [ X^J , X^K] \Big)  \label{grass}
\end{eqnarray}
Notice that (\ref{vacua}) is a set of matrix equations, while (\ref{grass}) is just a set of scalar equations. The set of solutions divides into two branches: on one branch $X_+^I X_+^I = 0$, on the other branch $X_+^I X_+^I > 0$.

If $X_+^I X_+^I > 0$ we can use $SO(8)$ rotations to set $X^I_+=(X^1_+,0,\cdots,0)$. It is then easy to see that the (\ref{vacua}) imply $[X^I,X^J]=0$ for any $I,J = \{ 1 \ldots 8 \}$, so that (\ref{grass}) is automatically satisfied. In the case $\mathcal{G} = SU(N)$, the solutions of (\ref{vacua}) and (\ref{grass}) are of the form
\begin{equation}
X^I = U^{-1} \big( D^I + M X^I_+ \big) U
\end{equation}
where $M$ is an adjoint matrix and $U$ is unitary. $X^I_-$ is unconstrained. We can at this point quotient by global gauge transformations. Setting $M = - D^1 / X^1_+$ we can have $X^1 = 0$, and also $X_-^1$ is fixed. The quotient by the compact part set $X^I$ to be diagonal and non vanishing for $I = 2 \ldots 8$. So it appears that the moduli space has dimension $7 (N - 1) + 16$. 
Dualizing the low energy gauge fields into scalars and keeping track of the shift symmetry of $X^I_-$,  this abelian branch seems to become
\begin{equation}
\label{branch1}
\mathcal{M}=\frac{(\mathbb{R}^8)^{N}}{S_N}\ ,
\end{equation}
where $S_N$ is the discrete group of permutations of the eigenvalues of the $X^I$ matrices. Note that (\ref{branch1}) is precisely the moduli space which one would expect for M2 branes in flat space.

On the other branch $X^I_+ = 0$, the eqs. (\ref{vacua}) are trivially satisfied, while the equations (\ref{grass}) impose a set of at most $\mathcal{C}\left( 8, 3 \right) = 56$ independent constraints on the $8$ matrices $X^I$. The non compact part of the gauge symmetry can be used to gauge fix $X^I_-$, thus we are led to the set of solutions to ${\rm Tr}\big( X^I [ X^J , X^K] \big) = 0$ modulo $\mathcal{G}$ transformations. The dimension of this branch is of order $N^2$.

\subsubsection{Massive excitations on the abelian branch}

Let us now examine the effective theory at a generic point on the moduli space $\bar{X}^I$, where $[\bar{X}^I,\bar{X}^J]=0$. We can take
\begin{equation}
\bar{X}^I={\rm diag}(a_1^I,\cdots,a^I_N)=a^I(\alpha)\,\delta^{\alpha}_{\beta}\, ,\quad \sum_{\alpha} a^I(\alpha)=0\ ;
\end{equation}
and then consider linearized fluctuations around this vacuum $X^I=\bar{X}^I+\epsilon^I$. Upon defining
\begin{equation}
\label{L^I}
L^I(\alpha,\beta)=a^I(\alpha)-a^I(\beta)\ ,
\end{equation}
it is easy to see that the linearized equations of motion for $\epsilon^I$ become
\begin{equation}
\partial^2(\epsilon^I)^{\alpha}_{\beta}-[M^2(\alpha,\beta)]^I_J(\epsilon^J)^{\alpha}_{\beta}=0\ .
\end{equation}
The mass-squared matrix, in $SO(8)$ space, reads
\begin{equation}
[M^2]^I_J=\Big(\vec{L}^2\,\vec{X}_+^2-(\vec{X}_+\cdot\vec{L})^2\Big)\delta^I_J-\vec{L}^2\,X_+^IX_+^J-\vec{X}_+^2\,L^IL^J+(\vec{X}_+\cdot\vec{L})\Big(X_ +^I L^J+X_+^JL^I\Big)\ ,
\end{equation}
where $L^I=L^I(\alpha,\beta)$. It is clear that, for $\alpha=\beta$, since $L^I=0$, the $(\epsilon^{I})^{\alpha}_{\alpha}$ will remain massless. However, the shift symmetry $X^I\rightarrow X^I+ M\, X_+^I$ allows to gauge-fix one of the $X^I$ directions. Therefore, the massless fields will be the $(\epsilon^I)^{\alpha}_{\alpha}$, with $I=1,\dots,7$. Given that the matrix $\epsilon^I$ has to be traceless, we are left with $7(N-1)$ real scalar fields parametrizing the moduli space. In addition, the from the covariant derivatives expanded around this vacuum, it is easy to see that the gauge field $B_{\mu}$ acquires a mass given by $(X_ +^I)^2$, while the $SU(N)$ part, corresponding to $A_{\mu}$, gets broken to $U(1)^{N-1}$. We can now dualize these $N-1$ abelian gauge fields to obtain $N-1$ scalars, in such a way that we are left with the $(\mathbb{R}^8)^N/S_N$. On the other hand, hadn't we dualized the abelian gauge fields, we would have $7(N-1)$ real scalars and $N-1$  gauge fields, which is the expected field content for D2 branes in flat space. In this case, for a large value of $X_+^I$, as shown in \cite{Gran:2008vi} along the lines of \cite{Mukhi:2008ux}, the theory precisely approaches the SYM in 3 dimensions corresponding to a stack of D2 branes. 

The massive excitations should carry information about the degrees of freedom through which the membranes interact. With no loss of generality we can choose $X_0^I$ to be aligned with the direction 1, \textit{i.e.} $X^I_0=(X^1_0,\cdots,0)$. Then, the mass-squared matrix can be easily diagonalized. The squared masses are 
\begin{equation}
m^2=(X_+^1)^2 (\vec{L}^2-(L^1)^2)\ ,
\end{equation}
in such a way that $m=X_+^1\sqrt{(L^2)^2+\cdots (L^8)^2}$. Note that the $L^1$ does not appear in the mass expression, which is, as expected, an area. However, at this point it is unclear the meaning of this area, which, however, seems to have one of its vertices trapped at $X_+^1$.

\section{Comments}\label{concl}

In this letter we proposed a large set of superconformal gauge theories described by a lagrangian with $\mathcal{N}=8$ supersymmetry which fits into the class of models introduced by Bagger, Lambert and Gustavsson. Part of the field content transforms in the adjoint of an $SU(N)$ gauge symmetry, which appears in the lagrangian through a $BF$ theory. The theory has a non-compact gauge symmetry, and allows for a moduli space consistent with a potential interpretation of the theory in terms of M2 branes in flat space-time. In further support of this interpretation, we have shown that the would-be coupling can always be reabsorbed via field redefinitions, in such a way that the theory does not really have a tunable coupling.


It is not evident that the theory is unitary, since some matter fields appear with the wrong sign in the lagrangian. We argued that the negative norm states can be eliminated via a suitable projection onto physical states. However, a more careful analysis should be undertaken. A related issue is the proper treatment of the shift symmetry of the fields $X^I_-$.


 \section*{Acknowledgments}
We would like to thank Juan Maldacena and Nathan Seiberg for important suggestions. We are also grateful to Marcus Benna, Seba Franco, Michele Papucci and Antonello Scardicchio for  discussions. E.T. is grateful to Princeton University for the warm hospitality during the development of this paper. D. R-G. acknowledges financial support from the European Commission through Marie Curie OIF grant contract no. MOIF-CT-2006-38381.

\appendix
\section{Check of the fundamental identity}\label{FI appendix}

In this appendix we check that the fundamental identity (\ref{FI}) is satisfied by the choice (\ref{basic3A}) for the structure constants $f^{abc}\,_d$. \\
As for the free low index $d$, when $d=+$ all the 4 terms in the l.h.s. of (\ref{FI}) vanish, therefore we are left with $d=-$ and $d=\delta$. From (\ref{basic3A}), the $-$ cannot appear among the upper indices and the low index cannot be $+$ to give non trivial result, therefore fro the repeated index we have $g=\rho$.\\
Starting with $d=-$, we get
\begin{equation}
\label{FI minus}
f^{ef\rho}\,_- f^{abc}\,_\rho
-f^{efa}\,_\rho f^{bc\rho}\,_-
-f^{efb}\,_\rho f^{ca\rho}\,_-
-f^{efc}\,_\rho f^{ab\rho}\,_-=0\ .
\end{equation}
When either $e=+$ or $f=+$ and the remaining indices are greek, the (\ref{FI minus}) reduces to the Jacobi identity
\begin{equation}
\label{Jacobi}
f^{\phi \alpha}\,_\rho \,f^{\beta\gamma\rho}
+f^{\phi\beta}\,_\rho \,f^{\gamma\alpha\rho} 
+f^{\phi\gamma}\,_\rho \,f^{\alpha\beta\rho} 
=0\ .
\end{equation}
Instead, when one index among $\{a,b,c\}$ is $+$ and all the other ones (included also $e$ and $f$) are greek, the two remaining terms in  the l.h.s. of (\ref{FI}) cancel between themselves. In all the other cases with $d=-$, the four terms in the r.h.s. of (\ref{FI}) simply vanish.\\
When $d=\delta$, then only one index $+$ must occur in the corresponding $f$
\begin{equation}
\label{FI delta}
f^{ef\rho}\,_\delta f^{abc}\,_\rho
-f^{efa}\,_\rho f^{bc\rho}\,_\delta
-f^{efb}\,_\rho f^{ca\rho}\,_\delta
-f^{efc}\,_\rho f^{ab\rho}\,_\delta=0\ .
\end{equation}

\end{document}